# Fast-Light in a Photorefractive Crystal for Gravitational Wave Detection

H.N. Yum[a], M. Salit[b], G.S. Pati[b], S. Tseng[b], P.R. Hemmer[a], and M.S. Shahriar[b]

[a] Department of Electrical Engineering, Texas A&M University,
College Station, Texas 77843

[b] Department of Electrical Engineering and Computer Science, Northwestern University,
Evanston, IL 60208

**Abstract:**

We demonstrate superluminal light propagation using two frequency multiplexed pump beams to produce a gain doublet in a photorefractive crystal of $Ce:BaTiO_3$. The two gain lines are obtained by two-wave mixing between a probe field and two individual pump fields. The angular frequencies of the pumps are symmetrically tuned from the frequency of the probe. The frequency difference between the pumps corresponds to the separation of the two gain lines; as it increases, the crystal gradually converts from normal dispersion without detuning to an anomalously dispersive medium. The time advance is measured as 0.28 sec for a pulse propagating through a medium with a 2Hz gain separation, compared to the same pulse propagating through empty space. We also demonstrate directly anomalous dispersion profile using a modfied experimental configuration. Finally, we discuss how anomalous dispersion produced this way in a faster photorefractive crystal (such as SPS: $Sn_2P_2S_6$) could be employed to enhance the sensitivity-bandwidth product of a LIGO type gravitational wave detector augmented by a White Light Cavity.

## 1. Introduction

In a so called fast-light medium, the dispersion is anomalous over a limited bandwidth. In such a medium, the group velocity for a pulse made up of frequency components within this bandwidth can be greater than the free space velocity of light. We have been exploring a range of applications for such a medium[1,2,3,4,5,6,7]. These include enhancement of the sensitivity-bandwidth product of a LIGO-type gravitational wave detector, optical data buffering with a delay far exceeding the limit imposed by the delay-bandwidth product of a conventional cavity, a zero-area Sagnac ring laser gravitational wave detector with augmented strain sensitivity, and a super-sensitive ring laser gyroscope.

These applications are all based on so-called the White Light Cavity (WLC). A WLC is a cavity which resonates over a broader range of frequencies than ordinary empty cavities of equal length and finesse without a reduction in the cavity lifetime. As such, it can circumvent the tradeoff between the resonance bandwidth and the field build-up factor that ordinary cavities entail. A WLC also has the property that if the cavity length is moved away from the condition for empty cavity resonance, the frequency offset needed to restore the resonance is much larger than that for a conventional cavity, thus making it a more sensitive displacement and rotation sensor than an empty cavity can be.

A variety of approaches have been proposed and studied for realizing a white light cavity experimentally.[8,9,10] For example we have previously demonstrated one approach that uses a dispersive vapor medium within the cavity.[3] Specifically, the anomalous dispersion was produced by a rubidium vapor with bi-frequency pumped Raman gain. A WLC operating at the wavelength of the Rb-transition, however, is not suitable for many of the applications listed above. For example, in order to apply the WLC concept for enhancing the bandwidth-sensitivity product of a LIGO-like gravtitational wave detector, it is necessary to realize a WLC that operates at 1064 nm.

In this paper, we show that the two wave mixing between pump and probe pulses in a photorefractive crystal creates a double gain profile similar to that of the bi-frequency pumped Raman gain in rubidium we have used previously, with the corresponding anomalous dispersion. We demonstrate superluminal propagation of pulses in such a medium. The anomalous dispersion produced in this way can also be employed to realize a WLC. The experiment reported here used a green laser at 532 nm, with a photorefractive crystal that has a relative slow (~seconds) response time. However, the technique demonstrated here is generic enough so that it can be employed with other photorefractive crystals as well. For example, a crystal of SPS [11,12] is sensitive at 1064 nm, and has a much faster (~msec) response time. An extension of this technique to the SPS crystal at 1064nm could thus be used to make a WLC for enhancing the sensitivity-bandwidth product of a LIGO-type gravitational wave detector as mentioned above.

*2. Gain Doublet in photorefractive regime*

The gain doublet and corresponding anomalous dispersion are the product of non-degenerate two-wave mixing and angular multiplexing in a photorefractive crystal. Here, we summarize briefly the physical model used to study this process. A space charge field is generated by the interference of two strong pump beams with a weak probe, so that the refraction index is modulated by the electro-optic effect. The probe is coupled with the pump beams due to these refractive index gratings[13]. Assuming that the intensity of each pump is much higher than that of the probe, in undepleted pumps approximation the phase and intensity coupling coefficients can be written as functions of the angular frequencies of the pumps and the probe[14]:

$$\Gamma_{in} = \sum_{j=1,2} \frac{\Gamma_{0j}}{2} \left[ \frac{d}{1+(\omega_s - \omega_{Pj})^2 \tau_j^2} \right] \quad (1a)$$

$$\Gamma_{ph} = \sum_{j=1,2} \frac{\Gamma_{0j}}{2} \left[ \frac{d(\omega_s - \omega_{Pj})\tau_j}{1+(\omega_s - \omega_{Pj})^2 \tau_j^2} \right] \quad (1b)$$

where $\Gamma_{in}$ and $\Gamma_{ph}$ are the intensity and phase coupling coefficients respectively. $\Gamma_{0j}$ depends on the incident angle between the jth pump and the probe[15]. $\tau_j$ is the rise time of the space charge field induced by the pumps, and d is the effective interaction length. $\omega_s$ and $\omega_{pj}$ are the angular frequencies of the probe and the j$^{th}$ pump respectively. The intensity gain and the phase shift of the probe are determined by these coefficients. For our experiment, the 1$^{st}$ and 2$^{nd}$ pump beams are up-shifted and down-shifted respectively by $\Delta\omega$ from the source frequency of $\omega_0$, so that $\omega_{p1}=\omega_0+\Delta\omega$, $\omega_{p2}=\omega_0-\Delta\omega$. This creates two gain lines separated by $2\Delta\omega$, and a region of anomalous dispersion occurs between them. The probe is pulsed, with a Gaussian spectrum centered around $\omega_0$.

The group velocity of the Gaussian pulse can be expressed as $v_g = C/(n + C \times \partial\Gamma_{ph}/\partial\omega)^{12}$, where C is the speed of light in vacuum and n is the mean index. If the carrier frequency of the probe pulse is placed in the middle of the anomalous dispersion region and the pulse bandwidth is smaller than the anomalous dispersion bandwidth, the group velocity in the photorefractive medium can become larger than *C*, or may even become negative. It is instructive to view these parameters graphically. We consider the probe pulse to be of the form $\exp(-t^2/t_0^2)$. We choose the pump intensities to be equal so that $\Gamma_{01} = \Gamma_{02} \equiv \Gamma_0$ and $\tau_1 = \tau_2 \equiv \tau_M$. For illustration, we consider $\Gamma_0 d = 6$, $\tau_M = 1.1\,\text{sec}$, and $t_0 = 0.6\,\text{sec}$. Figs. 1(a) shows the normalized intensity and phase coupling coefficients ($\Gamma_{in}$, $\Gamma_{ph}$), and the Fourier Transform (S) of the probe as functions of $\omega_s$- $\omega_0$, for $\Delta\omega$=0. Here the two gains overlap exactly and behave like a single gain line. The dispersion around the probe carrier frequency in this case is normal, and the v$_g$ of the probe becomes smaller than *C*. The width of the gain here is close to $\tau_M^{-1}$. In Figs 1(b-d), we plot the same parameters for increasing separation between the pump frequencies. For $\Delta\omega = 1$Hz (Fig. 1(c)), the separation is comparable to the gain width, so that two gain peaks are clearly distinguishable. This leads to anomalous dispersion between the peaks, as shown. In Fig(d), the separation becomes sufficiently wide so that that anomalous region covers the whole bandwidth of the Gaussian spectrum of the pulse.

### *3. Pulse propagation in the gain doublet system of photorefractive materials*

For the case of the Gaussian input pulse of the form $\exp(-t^2/t_0^2)$, the output pulse coupled in the frequency domain can be expressed as[11,14]

$$S(d, \omega_s) = S(0, \omega_s) \exp(\Gamma_{in1} + \Gamma_{in2}) \exp(i(\Gamma_{ph1} + \Gamma_{ph2})) \quad (2)$$

where d is the propagation distance of the probe inside a photorefractive crystal. $S(0, \omega_s)$ is the Fourier transform of the input pulse at the entrance of the material. Hence, one can obtain the total output signal intensity in the time domain by squaring the inverse Fourier transform of $S(d, \omega_s)$. Figs. 2(a)~(d) show the numerical simulation of the normalized output signal intensity for different gain line separations corresponding to each case in Figs. 1(a)~(d). In Figs. 2(a) the output is clearly delayed compared to the pulse propagating in free space, as expected. Fig. 2(b) shows the resultant output from propagation under the two gain peaks configuration associated with Fig. 1(b). To understand the behavior of the probe in this case, note brief that the gain in Fig. 1(b) is peaked at $\omega_0 \pm \Delta\omega$ with a deep valley in the middle. As such, the components of the probe spectrum at $\omega_0 \pm \Delta\omega$ get amplified disproportionately, leading to a two peaked spectrum. This leads to the beat note at 1Hz in Fig. 2(b). Furthermore, each of these peaks experiences normal dispersion, which leads to slowing of the probe, also evident in Fig. 2(b). In the case of Fig. 1(c), the gain separation is large enough so that the spectrum of the probe is almost completely within the region where the dispersion is anomalous (i.e. negative). This leads to advancement of the probe pulse, as can be seen in Fig. 1(c). The advance in this case is rather small due to the moderate steepness in the anomalous dispersion, and is evident only near the peak. Note also that a residual beat note is present at 2Hz, as expected. As the gain separation increases further (Fig 1(d)), the slope of the anomalous dispersion become smaller, thus reducing the pulse advancement, as seen in Fig. 2(d). It is evident from Figs. 1(c) and 1(d) that $\partial^2 \Gamma_{ph}/\partial \omega^2 \neq 0$ for $\omega \neq \omega_s$. The resulting group velocity dispersion [16,17] causes pulse compression in this case , as can be seen clearly in Figs. 2(c) and 2(d).

### *4. Experimental pulse propagation set-up and result*

We carried out an experiment that corresponds closely to the simulation. The experimental set-up is illustrated schematically in Fig.3. A collimated 532nm doubled Nd-YAG laser was split into a probe and two pump beams. The acousto-optic modulators

(AOMs) were driven by frequency synthesizers (PTS's). The IQ modulator (JCIQ-176M, Minicircuit) and the AOM generated the Gaussian probe pulse with a carrier frequency of $f_L$+110MHz+4Hz corresponding to $\omega_0/2\pi$, where $f_L$ is the laser frequency. The Gaussian pulse with a temporal width of $t_0$=0.6sec was generated by a DAQ-Card (DAQCard- 6036E, National instrument) with a repetition period of 20sec for $\Delta\omega/2\pi$=0 and 0.5Hz and, 10sec for $\Delta\omega/2\pi$=1 and 2Hz. The first and second pumps (P1 and P2) were shifted by $\pm\Delta\omega/2\pi$ from $f_L$+110MHz+4Hz, respectively, thereby producing two gain peaks with a separation of $2\Delta\omega/2\pi$. The incident angle of the probe is approximately 90° from the C-axis of the Ce:BaTiO$_3$ crystal used for this experiment. P1 and P2 were angular multiplexed at angles of 40° and 60° with respect to the probe. The probe was coupled to the two pumps over an interaction length of 0.5cm inside the crystal. The polarization direction of all the beams and the C-axis were parallel to the optical table. A part of the probe was split-off to provide a reference pulse. This probe was monitored simultaneously with the pulse that traveled through the crystal, and was used to determine the degree of pulse delay/advancement and compression.

Figs. 4(a)~(d) show the normalized reference and output signals as $\Delta\omega$ increases. For $\Delta\omega$=0 in Fig. 4(a), the two gains coincide to form a single gain which results in the time delay (1.3sec) of the Gaussian input pulse. As $\Delta\omega/2\pi$ increases to 0.5Hz, parts of the probe spectrum shifted from $\omega_0$ by ±0.5Hz (the maximum gain and slowing region) become the primary frequency components which are amplified and delayed, thereby resulting in the beat frequency of 1Hz, corresponding to $2\Delta\omega/2\pi$, as shown in Fig. 4(b). In Fig. 4(c), since the gain doublet is sufficiently separated so that the spectrum of the probe exists entirely within the anomalous dispersion region, the output is advanced by 0.28sec compared to the reference. As $\Delta\omega/2\pi$ increases up to 2Hz, the anomalous dispersion becomes very small. Hence, it is observed that the output has virtually no advancement, as can be seen in Fig. 4 (d).

**5. Experimental dispersion measurement set-up and result**

In these pulse propagation experiments, the dispersion was not measured directly, but rather inferred from the gain conditions and the group velocity. To measure the dispersion directly, we used a different experimental configuration, as shown in figure 5, top. Briefly, the output of the frequency-double Nd:YAG laser at 532 nm was split in two parts. One part was frequency shifted by an AOM, operating at 40 MHz. The other part remained unshifted. The unshifted beam was split further into two parts: one denoted as pump 1 and the other as signal 1. Similarly, the shifted beam was split further into two parts: pump 2 and signal 2. Pump 2 (signal 2) was combined with pump 1 (signal 1) with a beamsplitter. The combined beam containing both pumps was applied to the crystal directly. The combined beam containing both signals was reflected by a mirror mounted on a PZT (Piezoelectric Transducer) before being applied to the crystal, with a relative angle of about fifteen degrees with respect to the combined-pumps beam.

A voltage varying quadratically with time was applied to the PZT. This moved the mirror with a velocity that varied linearly with time. As such, each of the signal beams reflecting off this mirror experienced a Doppler effect induced frequency shift which also varied linearly with time. Thus, the effect of the PZT scan was to tune the frequency of signal 1 (signal 2) around the frequency of pump 1 (pump 2). The profile of the scanning voltage was chosen to be so that the frequency difference between the signal and the pump varied from -0.5 Hz to 0.5 Hz.

Consider first the experiment when signal 2 was blocked, but pump 1, pump 2 and signal 1 were on. The dispersion experienced by signal 1 due to the presence of pump 1 was measured using the following approach. A 1 kHz modulation was added to the scanning voltage applied to the PZT, using the local oscillator from a lock-in amplifier. This produced sidebands which interfered with a part of the pump diffracted into the direction of the probe due to the induced grating, producing a beat note. The phase of this beat signal depends upon the phase of the probe beam, and thus upon the index of refraction it experiences while passing through the crystal. The beat note was demodulated with a reference signal, producing an output proportional to the phase of the probe beam. Scanning the frequency of the probe (via the quadratic voltage applied to the PZT) while recording this signal allowed us to map out the index of refraction of the material as a function of frequency.

The dispersion observed using this method is shown in the left trace in the bottom part of figure 5. As expected, the dispersion is normal at the center, and anomalous on the sides, corresponding to the gain of signal 1. Even though pump 2 is present in the system, the dispersion as seen by signal 1 here is due primarily to the effect of pump 1. The experiment was then repeated with signal 1 blocked, but pump 1, pump 2 and signal 2 were on. The resulting dispersion, due to the gain of signal 2 caused primarily by pump 2, is shown in the right trace in the bottom part of figure 5. The two traces, taken together, and separated by 40 MHz, can be interpreted to be the dispersion seen by a single signal beam when its frequency is scanner over the whole range. If pump 1 and pump 2 were closer together (e.g., with a frequency separation of 2 Hz), then the right edge of the left trace would merge into the left edge of the right trace, producing a linear but anomalous dispersion for a signal beam with its frequency in-between those of pump 1 and pump 2. For technical constraints, we were not able to carry out this particular experiment with the frequencies of the two pumps so close to each other. In the experiment described in figure 3, the two pumps were angularly multiplexed. In contrast, in the experiment described in figure 5, the two pumps were superimposed on each other spatially. The latter approach may be more convenient for some applications.

As mentioned earlier, the anomalous dispersion demonstrated using a photorefractive crystals in these two experiments implies that it should be possible to realize a White Light Cavity using such a medium. For application to gravitational wave detection, the wavelength of interest is 1064 nm, since it is used by the current embodiment of the LIGO interferometer, and high quality mirrors and detectors suited for LIGO have been developed at this wavelength. A crystal of SPS ($Sn_2P_2S_6$) is sensitive at this wavelength, and could be employed for this purpose. Furthermore, an SPS crystal has a much faster response time (~msec)[11,12], which in turn means that the bandwidth of the negative dispersion regime can extend over a few kHz, which is optimal for detection of gravitational waves[5,7].

*5. Conclusion*

We have demonstrated the realization of negative dispersion and superluminal light propagation using a dual-frequency pumped photorefractive crystals. The results show that that such a medium is a promising candidate for the design of a White light Cavity (WLC). We have previously demonstrated a WLC with the transmission bandwidth of approximately 10 MHz, using the anomalous dispersion between the Raman gain peaks of $^{85}$Rb atomic vapor. The desired bandwidth for LIGO applications is only a few kHz. The bandwidth of the crystal used in this experiment was on the order of 1Hz, which is too narrow for such an application. However, the bandwidth can be increased up to a few KHz by using a fast response material such as an SPS crystal.

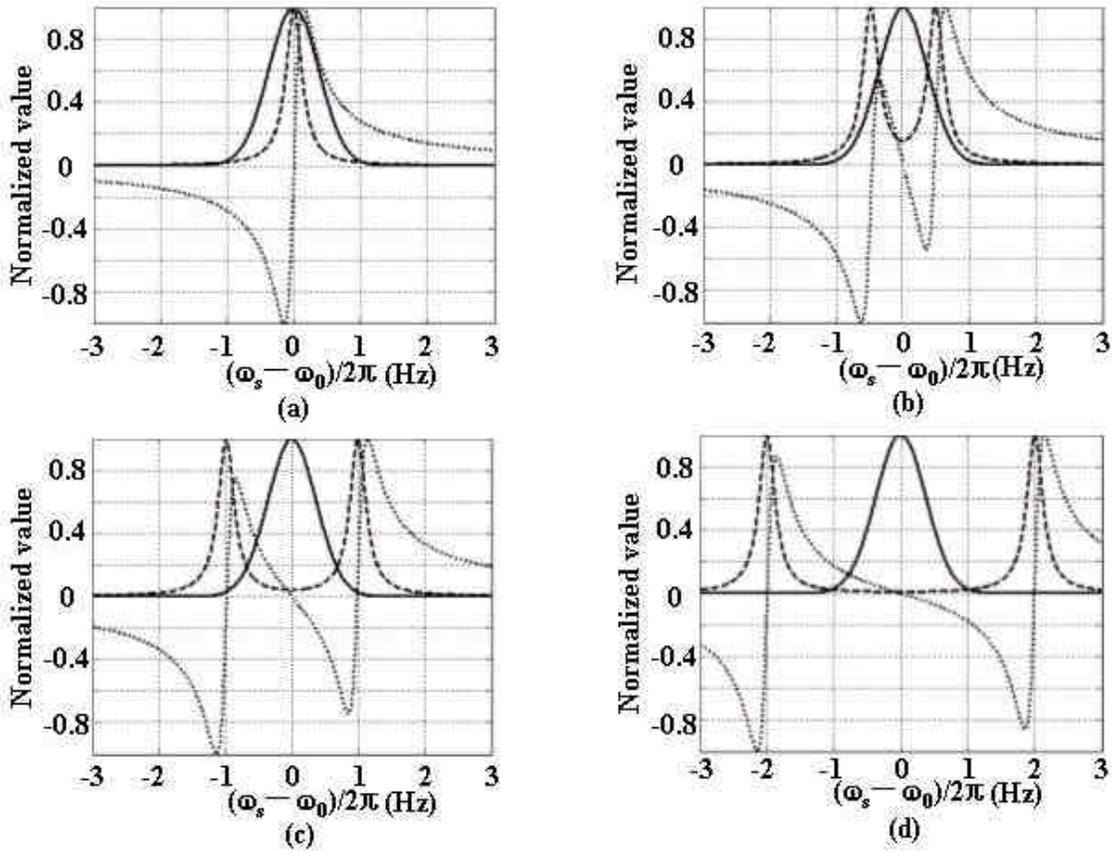

Fig. 1. Numerical, normalized plots of S(ω), the Fourier transform of the input pulse (solid line), $\Gamma_{in}$(dashed line), and $\Gamma_{ph}$(dotted line). The input pulse is assumed to be of the form $\exp(-t^2/t_0^2)$. For these plots, we have used for $t_0$=0.6sec, $\Gamma_0 d = 6$ and $\tau_1 = \tau_2 \equiv \tau_M = 1.1$sec. The four sets are for four different gain separations, 2Δω (a) 0Hz, (b) 1Hz, (c) 2Hz, (d) 4Hz

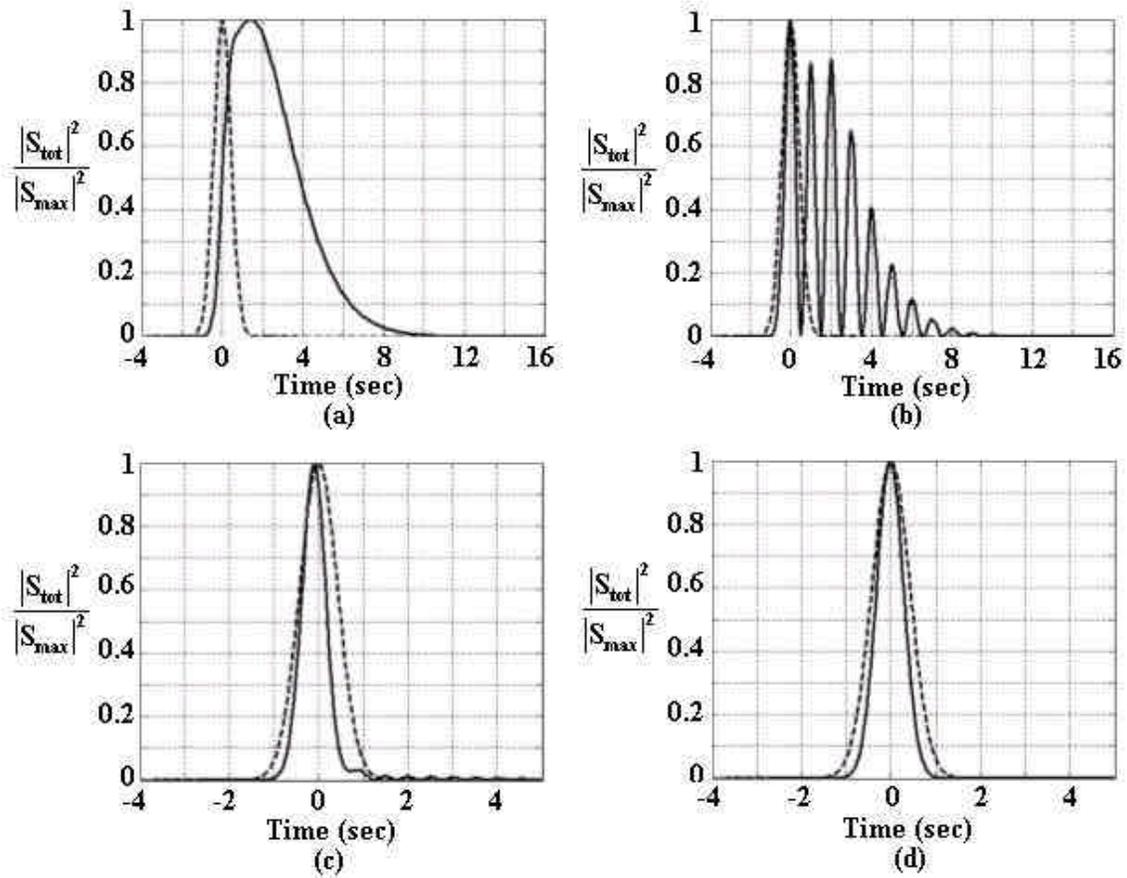

Fig. 2. Simulated output signal for the Gaussian input of $\exp(-t^2/t_0^2)$ for the same parameters as set in Fig. 1. The frequency difference between the gains is given as (a) 0Hz, (b) 1Hz, (c) 2Hz, and (d) 4H. Dashed line indicates reference, solid line indicates signal output.

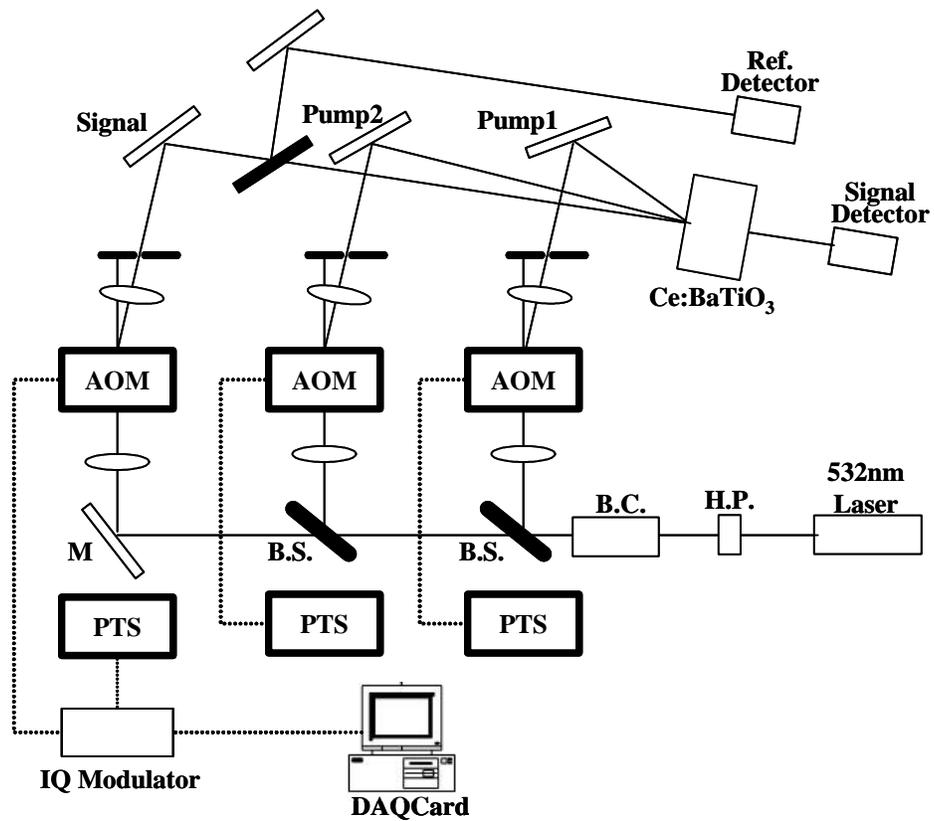

Fig.3. Schematic diagram of the experimental set-up; AOM : Acousto-optic modulator, B.S : Beam Splitter, B.C : Beam Collimator, H.P: Half waveplate, M: Mirror, PTS : Frequency synthesizer, Pump1: $f_L+110MHz+4Hz+\Delta\omega$, Pump2: $f_L+110MHz+4Hz-\Delta\omega$, Probe : Gaussian pulse with the carrier frequency of $f_L+110MHz+4Hz$

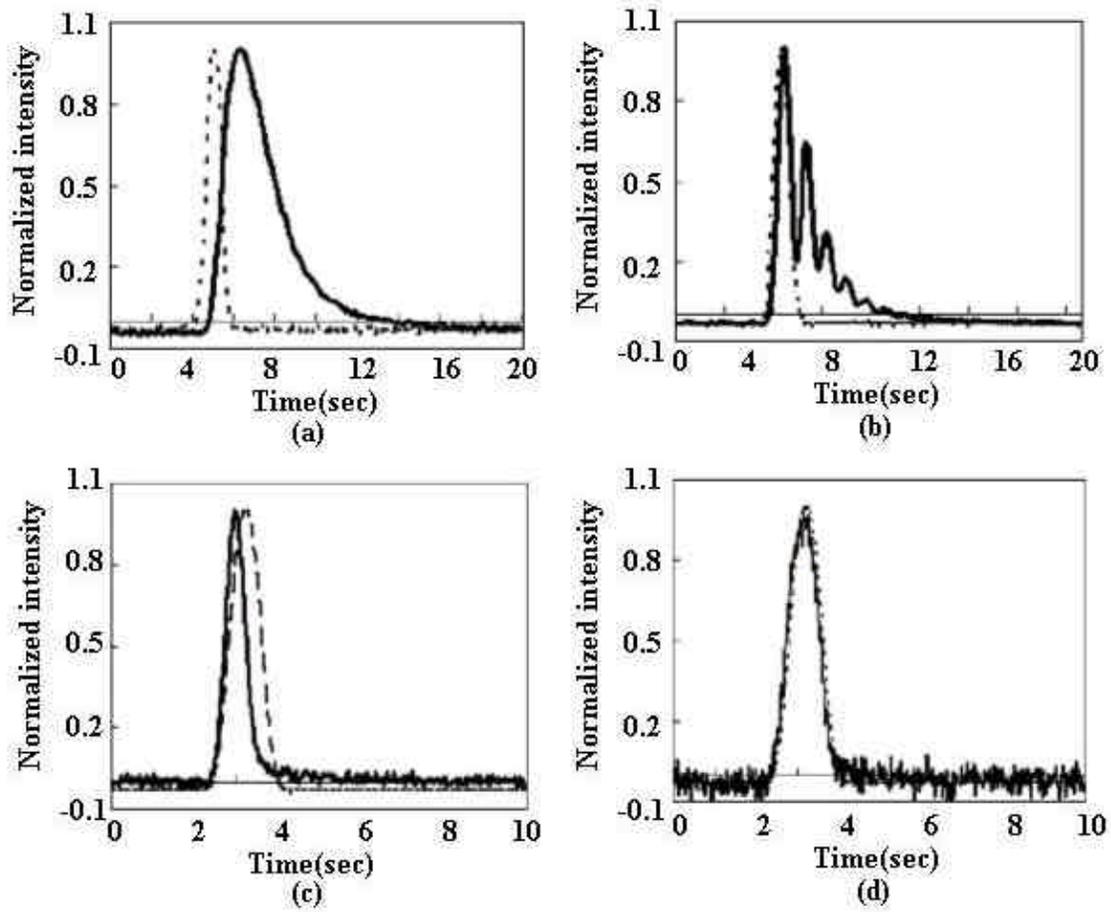

Fig. 4. Experimental results showing the group velocity variation of the signal output with setting the separation of the gain doublet to (a) 0Hz, (b) 1Hz, (c) 2Hz, and (d) 4Hz. Dashed line indicates reference, solid line indicates signal output

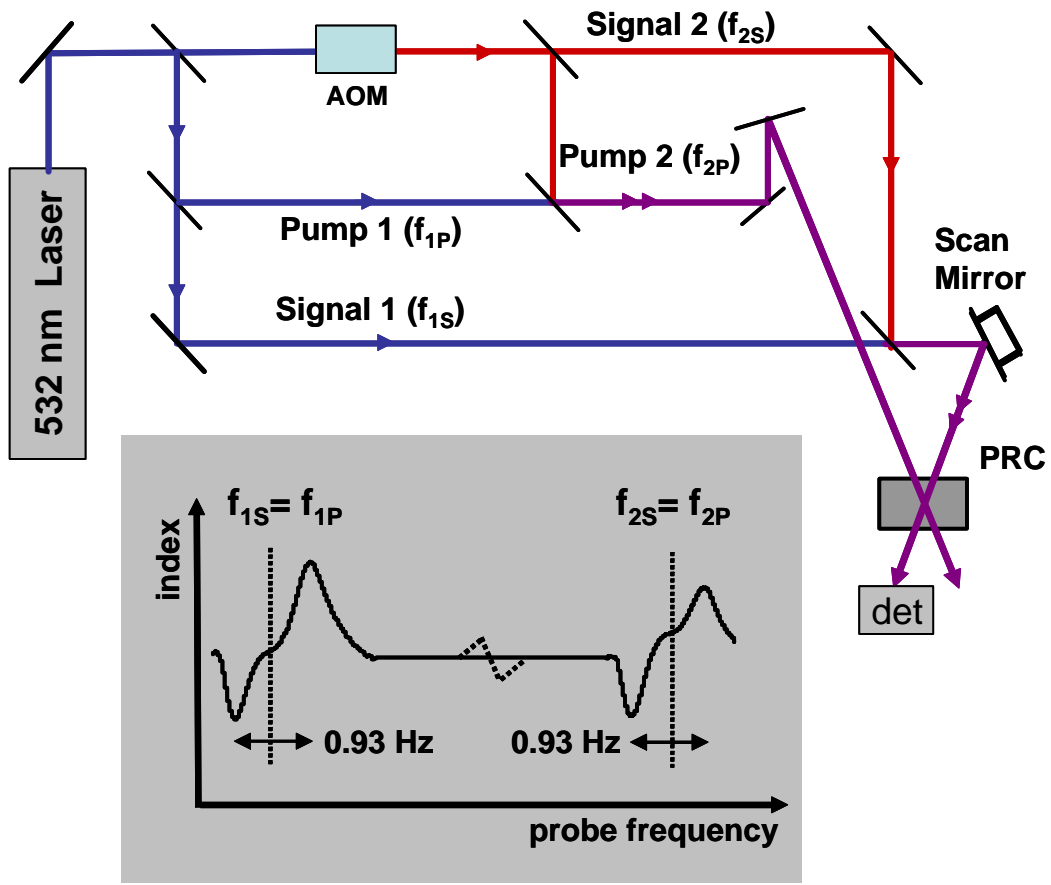

Fig. 5. (Top) Experimental set-up for measuring the dispersion profile directly. (Bottom) Observed dispersion profile centered around two pump frequencies. See text for additional details.

[1] M.S. Shahriar, G.S. Pati, R. Tripathi, V. Gopal, M. Messall and K. Salit, "Ultrahigh Precision Absolute and Relative Rotation Sensing using Fast and Slow Light", Physical Review A 75 (5): Art. No. 053807 MAY 2007.

[2] G.S. Pati, M. Messall, K. Salit, and M.S. Shahriar, "Demonstration of Tunable Displacement-Measurement-Sensitivity using Variable Group Index in a Ring Resonator," Optics Communications, 281 (19), p.4931-4935, (2008).

[3] G.S. Pati, M. Messall, K. Salit, and M.S. Shahriar, "Demonstration of a Tunable-Bandwidth White Light Interferometer using Anomalous Dispersion in Atomic Vapor," Phys. Rev. Lett. 99, 133601 (2007).

[4] M. Salit, G.S. Pati, K. Salit, and M.S. Shahriar, "Fast-Light for Astrophysics: Super-sensitive Gyroscopes And Gravitational Wave Detectors," Journal of Modern Optics, Volume 54, Issue 16 & 17, pages 2425 – 2440, (2007).

[5] M.S. Shahriar and M. Salit, "Application of Fast Light in Gravitational Wave Detection with Interferometers and Resonators," Journal of Modern Optics (2008), In-Press.

[6] M.S. Shahriar and M. Salit, "A Fast-Light Enhanced Zero-Area Sagnac Ring Laser Gravitational Wave Detector," to be submitted to Phys. Rev. Letts. (http://lapt.eecs.northwestern.edu/preprints/FE-ZASRL-GWD.pdf)

[7] M. Salit and M.S. Shahriar, "Enhancement of Sensitivity-Bandwidth Product of Interferometric Gravitational Wave Detectors using White Light Cavities," submitted to Physical Review A.

[8] S. Wise V. Quetschke, A. J. Deshpande, G. Mueller, D. H. Reitze, D. B. Tanner, and B. F. Whiting , "Phase Effects in the Diffraction of Light: Beyond the Grating Equation," Phys. Rev. Lett. 95, 013901 (2005).

[9] A.A. Savchenkov, A.B. Matsko, L. Maleki, "White-light whispering gallery mode resonators," Opt. Lett. 31, 92 (2006)

[10] A. Wicht, K. Danzmann, M. Fleischhauer, M. Scully, G. Miiller, R.H. Rinkleff, Opt. Communications" White-light cavities, atomic phase coherence, and gravitational wave detectors," 134, 431, (1997).

[11] E. Podivilov, B. Sturman, A. Shumelyuk and S. Odoulov, "Light pulse slowing down up to 0.025cm/s by photorefractive two-wave coupling", Phys. Rev. Lett. 91, 083902 (2003)

[12] Shumelyuk, K. Shcherbin, S. Odoulov, B. Sturman, E. Podivilov, K. Buse, "Slowing down of light in photorefractive crystals with beam intensity coupling reduced to zero", Phys. Rev. Lett. 93, 243604 (2004)

[13] L. Solymer, D.J. Webb, and A. Grunnet-Jepsen, "The physics and application of photorefractive materials," Clarendon Press, Oxford, (1996)

[14] Z. Deng, De-Kui Qiing, P.R. Hemmer, C.H. Raymond Ooi, M.S. Zubairy, and M.O. Scully, "Time-bandwidth problem in room temperature slow light", Phys. Rev. Lett. 96, 023602 (2006)

[15] P. Yeh, "Two-wave mixing in nonlinear media," IEEE J. Quantum Electronics, 25, 484 (1989).

[16] A. Dogariu, A. Kuzmich, and L.J. Wang, "Transparent anomalous dispersion and superluminal light-pulse propagation at a negative group velocity", Phys. Rev. A, 63, 053806 (2001).

[17] G. Zhang, R. Dong, F. Bo, and J. Xu, "Slowdown of group velocity of light by means of phase coupling in photorefractive two-wave mixing", Appl. Opt. 43, 1167 (2004).